\documentclass[conference]{IEEEtran}
\IEEEoverridecommandlockouts
\usepackage{cite}
\usepackage{amsmath,amssymb,amsfonts}
\usepackage[linesnumbered,ruled]{algorithm2e}
\usepackage{algpseudocode}
\usepackage{graphicx}
\usepackage{textcomp}
\usepackage{bm}
\usepackage[tight,footnotesize]{subfigure}
\usepackage{multirow}
\usepackage{xcolor}
\usepackage[square,numbers,sort&compress]{natbib}

\def\BibTeX{{\rm B\kern-.05em{\sc i\kern-.025em b}\kern-.08em
    T\kern-.1667em\lower.7ex\hbox{E}\kern-.125emX}}

\begin{document}



\title{a-Tucker: Input-Adaptive and Matricization-Free Tucker Decomposition for Dense Tensors on CPUs and GPUs}


\author{\IEEEauthorblockN{
Min Li\IEEEauthorrefmark{1}\IEEEauthorrefmark{2},
Chuanfu Xiao\IEEEauthorrefmark{3},
Chao Yang\IEEEauthorrefmark{3}
}
\IEEEauthorblockA{\IEEEauthorrefmark{1}Institute of Software, Chinese Academy of Sciences, Beijing 100190, China}
\IEEEauthorblockA{\IEEEauthorrefmark{2}University of Chinese Academy of Sciences, Beijing 100049, China}
\IEEEauthorblockA{\IEEEauthorrefmark{3}Peking University, Beijing 100871, China}
\IEEEauthorblockA{Corresponding author: Chao Yang (chao\_yang@pku.edu.cn.).}
}

\maketitle

\begin{abstract}

Tucker decomposition is one of the most popular models for analyzing and compressing large-scale tensorial data.
Existing Tucker decomposition algorithms usually rely on a single solver to compute the factor matrices and core tensor,
and are not flexible enough to adapt with the diversities of the input data and the hardware.
Moreover, to exploit highly efficient GEMM kernels, most Tucker decomposition implementations make use of explicit matricizations,
which could introduce extra costs in terms of data conversion and memory usage.
In this paper, we present a-Tucker, a new framework for input-adaptive and matricization-free Tucker decomposition of dense tensors.
A mode-wise flexible Tucker decomposition algorithm is proposed to enable the switch of different solvers for the factor matrices and core tensor,
and a machine-learning adaptive solver selector is applied to automatically cope with the variations of both the input data and the hardware.
To further improve the performance and enhance the memory efficiency, we implement a-Tucker
in a fully matricization-free manner without any conversion between tensors and matrices.
Experiments with a variety of synthetic and real-world tensors show that a-Tucker can substantially outperform existing works on both CPUs and GPUs.

\end{abstract}


\section{Introduction}
Tensor decomposition is a powerful tool for analyzing and compressing large-scale tensorial structured data.
Among many tensor decomposition models \cite{Hitchcock1927,Levin1963,Lathauwer2000-1,Oseledets2011},
the Tucker decomposition \cite{Levin1963, tucker-decom}, also known as the higher-order singular value decomposition (HOSVD) \cite{Lathauwer2000-1},
is among the most popular ones.
In recent years, Tucker decomposition is gaining increasingly more attention and has been widely used
in scientific computing \cite{ipdps16-tuckermpi}, data mining \cite{healthcare,detect-anomalous-behavior},
and computer vision \cite{app-image-process-1,app-image-process-2,cnncompression} fields.
Therefore, it is of great importance to study high performance Tucker decomposition computation on modern hardware platforms.
In this paper, we aim to study high performance computation of dense Tucker decomposition based on the popular sequentially
truncated singular value decomposition ($st$-HOSVD) algorithm \cite{sthosvd-algo}.

During the computing of the $st$-HOSVD algorithm, two main steps are carried out to compute the factor matrices and update the core tensor.
The original solver for factor matrices and core tensor is based on the singular value decomposition (SVD) of the matricized tensor \cite{sthosvd-algo},
which can be replaced with, for example, the eigen-decomposition of the Gram matrix \cite{ipdps16-tuckermpi},
or an alternating least squares (ALS) method \cite{xiaoals}.
Based on these solvers, Tucker decomposition has been widely studied on high performance computers
\cite{tensortoolbox, tensorlab, ipdps16-tuckermpi, tuckermpi-software, sc18-dense-sthosvd-multi-gpu, csf-intro, hiCOO, b-csf, mixformat,sparse-ttm-gpu, algobenchmark}.
However, most of the works depend on a single solver for the factor matrices and core tensor,
and little attention has been paid to the effects of different solver variants.

Despite the continuing efforts made to improve the performance of $st$-HOSVD algorithm,
there are three major challenges arising from the application as well as the hardware development.
Firstly, more and more tensors with various dimensions and truncations are coming into usage in real-world applications.
It is observed that the existing approaches for computing factor matrices and core tensor are  only favorable for some certain input sizes,
and a single solver is usually not enough to achieve good performance for all input cases (see Figure 2).
Therefore, it is interesting to study how to design a flexible algorithm that can take different solvers into account.
In this case, the dynamic switch between different solvers is of great research and practical values.
Secondly, it is also worth studying how to select the most suitable solver according to the characteristic of the inputs,
especially on different hardware platforms.
The solver selector should be aware of the inputs and the hardware and at the same time yield relatively high accuracy and low overhead.
And thirdly, there are also challenges when implementing $st$-HOSVD algorithms on high performance platforms such as CPUs and GPUs.
To exploit highly efficient GEMM kernels, most conventional implementations rely on explicit matricization of the tensor to simplify the treatment of various tensor operations
\cite{tensortoolbox, tensorlab, xiaoals, sc18-dense-sthosvd-multi-gpu}, which, however, will bring more memory usage as well as extra conversion overheads,
and therefore leading to inferior performance.

To tackle these challenges, we propose a-Tucker, a new framework for input-adaptive and matricization-free Tucker decomposition of dense tensors.
A mode-wise flexible Tucker decomposition algorithm is proposed to enable the switch of different solvers for factor matrices and core tensor with
different tensor modes. To automatically select the most appropriate solver at runtime, a machine learning model is utilized to
help the flexible algorithm adapt with different inputs and also easily deploy on different hardware platforms.
Besides, matricization-free technique is employed to further improve the performance and reduce the memory usage.
Experiments show that a-Tucker can achieve substantially higher performance when compared with
the previous state-of-the-arts while keeping similar accuracy with both synthetic and real-world tensors.
As far as we are aware, this work is the first Tucker decomposition computing framework
which considers the effects of different solvers for factor matrices and core tensor and
can adapt with various sizes of the input tensor and the truncations on both CPUs and GPUs.

{The remainder of the paper is organized as follows.
In Section \ref{sec-background}, we introduce the background and motivation of this work.
We then present a highly flexible $st$-HOSVD algorithm and a corresponding adaptive solver selector
in Section \ref{algo} and Section \ref{adapt}, respectively.
Following that, implementation and optimization details are provided in Section \ref{opt}.
We give performance evaluations based on a variety of test tensors in Section \ref{experiment}
and briefly mention some related works in Section \ref{related-work}.
And the paper is concluded in Section \ref{conclusion}.}


\section{Background and Motivation}
\label{sec-background}
\subsection{Tensor notation and operations}

In this work we denote tensors, also known as multi-dimensional arrays, as boldface Euler script letters, e.g., $\bm{\mathcal{X}}$.
The order of a tensor is the number of dimensions, also called as modes of the tensor.
For example, an $N$th-order tensor can be expressed as $\bm{\mathcal{X}} \in\mathbb{R}^{I_{1}\times I_{2}\times\cdots\times I_{N}}$,
where $I_n$ denotes the {dimension} of mode-$n$.
In particular, vectors and matrices are first-order and second-order tensors
that are denoted by boldface lowercase and capital letters such as $\bm{x}$ and $\bm{X}$, respectively.
The elements of a tensor are scalars, and we denote them by lowercase letters with subscripts,
such as $x_{ijk}$ for the $(i,j,k)$ element of a third-order tensor $\bm{\mathcal{X}}$.
Analogous to matrices, tensors can be stored in row-major or column-major layouts.
For example, when using the column-major layout, elements in the first dimension of the tensor changes fastest.


The mode-$n$ matricization, also called unfolding operation,
reorders the elements of a tensor to form a matrix with the $n$-th dimension being the leading dimension.
For example, the mode-$n$ matricization of $\bm{\mathcal{X}} \in\mathbb{R}^{I_{1}\times I_{2}\times\cdots\times I_{N}}$
is $\bm{X_{(n)}}\in\mathbb{R}^{I_n \times J_n}$, where $J_n = I_{1} \cdots I_{n-1}  I_{n+1} \cdots  I_{N}$.
The Frobenius norm of tensor $\bm{\mathcal{X}}$ is defined as
\begin{align}
\|\bm{\mathcal{X}}\|_{F} = \sqrt{\sum\limits_{i_{1},i_{2},\cdots,i_{N}}{x}_{i_{1}i_{2} \cdots i_{N}}^{2}}.
 \label{equation-1}
\end{align}
It is easy to see that the Frobenius norm of a tensor is equivalent to the Frobenius norm
of the matricization of the tensor with respect to any mode.

The mode-$n$ tensor-times-matrix multiplication (TTM)
defines multiplications between a tensor $\bm{\mathcal{X}} \in\mathbb{R}^{I_{1}\times I_{2}\times\cdots\times I_{N}}$
and a matrix $\bm{U} \in\mathbb{R}^{R_{n}\times I_{n}}$, and produces
another tensor $\bm{\mathcal{Y}} = \bm{\mathcal{X}} \times_{n}\bm{U}$,
where $\bm{\mathcal{Y}} \in\mathbb{R}^{I_{1}\times\cdots\times I_{n-1}\times R_{n}\times I_{n+1}\times\cdots\times I_{N}}$.
The element-wise expression for TTM is as follows:
\begin{align}
y_{i_1 \cdots i_{n-1} j i_{n+1} \cdots  i_{N} }
&= (\bm{\mathcal{X}} \times_{n}\bm{U})_{i_1 \cdots i_{n-1} j i_{n+1} \cdots  i_{N}} \nonumber \\
&= \sum_{i_n=1}^{I_n} x_{i_1 i_{2} \cdots i_{N}}  u_{j i_{n}}.
 \label{equation-2}
\end{align}



Generally, the multiplication between two tensors is called tensor-times-tensor (TTT).
For example, let $\mathcal{I} = \mathcal{J} = \{1,\cdots,n-1,n+1,\cdots,N\}$ be two index sets,
the mode-$(\mathcal{I},\mathcal{J})$ product of two tensors $\bm{\mathcal{X}}\in\mathbb{R}^{I_{1}\times\cdots\times I_{n}\times\cdots\times I_{N}}$
and $\bm{\mathcal{Y}}\in\mathbb{R}^{J_{1}\times\cdots\times J_{n}\times\cdots\times J_{N}}$ with common modes $I_{m} = J_{m},\ m\neq n$,
can produce a second-order tensor $\bm{\mathcal{Z}}$, i.e., matrix, 
where its entries are given by
\begin{align}
z_{i_n r_n }
&= \sum\limits_{i_{1},\cdots,i_{n-1},i_{n+1},i_{N}=1}^{I_{1},\cdots,I_{n-1},I_{n+1},\cdots,I_{N}} x_{i_1 \cdots i_{n} \cdots i_{N}}  y_{j_1 \cdots r_{n} \cdots j_{N}}.
 \label{equation-3}
\end{align}


\subsection{Tucker decomposition}

\begin{figure}[!htb]
\centering
\includegraphics[width=0.72\columnwidth]{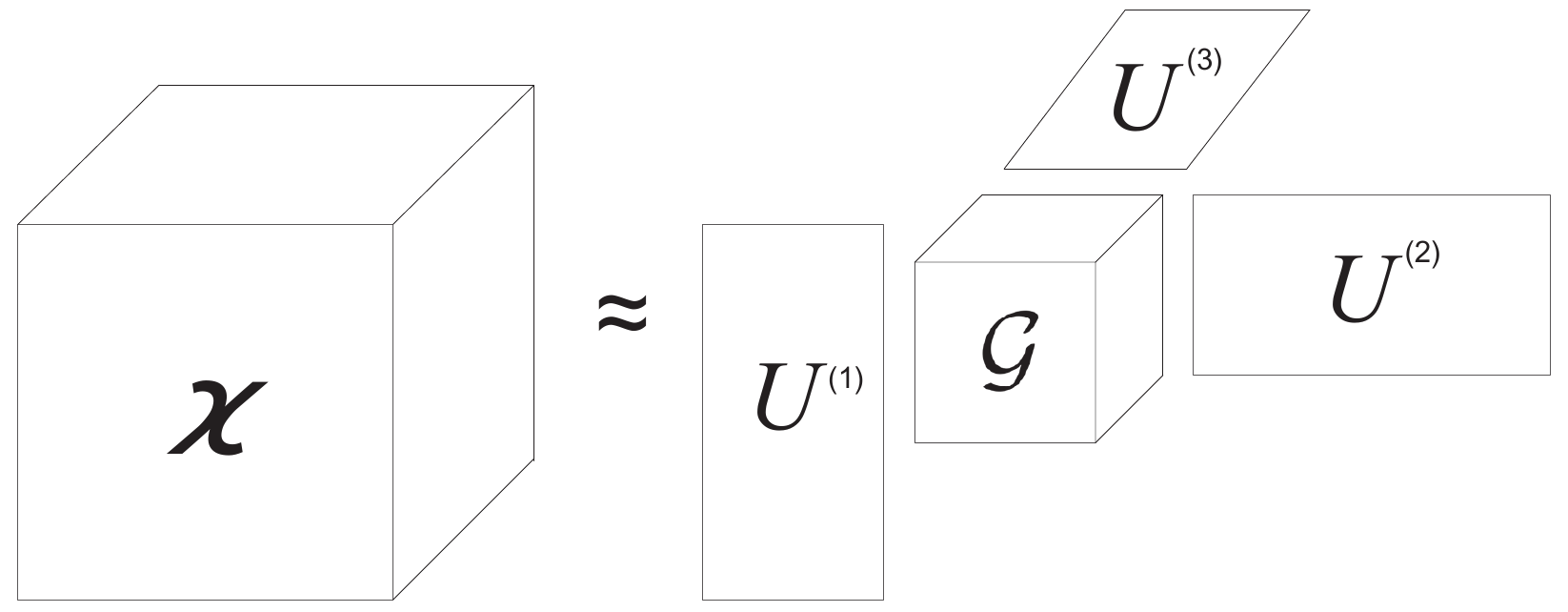}
\caption{Illustration of the Tucker decomposition for a third-order tensor. }
\label{fig-tensor-decom}
\end{figure}

The Tucker decomposition \cite{Levin1963, tucker-decom} approximates tensor $\bm{\mathcal{X}}$ of size ${I_{1}\times I_{2}\times\cdots\times I_{N}}$ as
$$ \bm{\mathcal{X}} \approx \bm{\mathcal{G}}\times_{1}\bm{U}^{(1)}\times_{2}\bm{U}^{(2)}\cdots\times_{N}\bm{U}^{(N)},$$
where $\bm{\mathcal{G}}$ is the core tensor, whose size is ${R_{1}\times R_{2}\times\cdots\times R_{N}}$ and
$\bm{U}^{(n)}$ is a factor matrix of size $I_n\times R_n$ for $n = 1, 2, \cdots, N.$
Usually the truncation $R_n$ is much smaller than $I_n$ due to the low-rank property of tensor decomposition.
An example of the Tucker decomposition  for a third-order tensor is illustrated in Figure \ref{fig-tensor-decom}.

The most common Tucker decomposition algorithms are the truncated higher-order singular value decomposition ($t$-HOSVD) \cite{Lathauwer2000-2, tensor-review}
and its sequentially truncated version ($st$-HOSVD) \cite{sthosvd-algo},
and the higher-order orthogonal iteration (HOOI) \cite{Lathauwer2000-2}.
The $st$-HOSVD or $t$-HOSVD algorithm is often used to initialize the factor matrices for HOOI.
For some applications, it has been shown that using $st$-HOSVD or $t$-HOSVD alone is in most cases sufficient to produce accurate results,
where HOOI only improves little accuracy \cite{sthosvd-algo}.
As compared to $t$-HOSVD, the $st$-HOSVD algorithm differs in shrinking the tensor in each mode
and reducing the subsequent computations {so as to reduce the computing cost and,
in most cases, improve the accuracy {\cite{sthosvd-algo}}.}
Therefore, in this paper we focus on high performance computation of the $st$-HOSVD algorithm.
We remark that, owning to the similar algorithm structure, the proposed ideas and optimizations can also be extended to other
Tucker decomposition algorithms, which is left as our future work.

\subsection{The $st$-HOSVD algorithm}

\begin{algorithm}
\label{algo-st-hosvd-eig}
\caption{The sequentially truncated higher-order singular value decomposition ($st$-HOSVD) algorithm.}
\SetAlgoLined
 \KwIn{Input Tensor $\bm{\mathcal{X}}\in\mathbb{R}^{I_{1}\times I_{2}\times\cdots\times I_{N}}$, \\
      Truncations $(R_{1},R_{2},\cdots, R_{N})$  }
\KwOut{Core tensor $\bm{\mathcal{G}}\in\mathbb{R}^{R_{1}\times R_{2}\times\cdots\times R_{N}}$, \\
       Factor matrices $\bm{U}^{(n)}\in\mathbb{R}^{I_{n}\times R_{n}}$ $(n=1,2,\cdots,N)$  }
$\bm{\mathcal{Y}} \leftarrow \bm{\mathcal{X}} $ \\
 \For{$n \leftarrow 1 \ \KwTo \ N $} {
      Mode-$n$ matricization $\bm{Y}_{(n)} \leftarrow  \bm{\mathcal{Y}} $ \\
      $ \triangleright$ {Compute factor matrices $\bm{U}^{(n)}$ }\\
      $\bm{U, \Sigma, V^T}  \leftarrow$ Singular value decomposition on $\bm{Y}_{(n)} $ \\
      $\bm{U}^{(n)} \leftarrow \bm{U}$ \\
      $ \triangleright$ Update core tensor $\bm{\mathcal{Y}}$ \\
      $\bm{Y}_{(n)} \leftarrow$ $\bm{\Sigma}{\bm{V^T}} $ \\
      Tensorization $\bm{\mathcal{Y}} \leftarrow  \bm{Y}_{(n)} $ \\
  }
 $\bm{\mathcal{G}} \leftarrow \bm{\mathcal{Y}}$
\end{algorithm}



As shown in Algorithm \ref{algo-st-hosvd-eig}, the computing procedure of the $st$-HOSVD algorithm is
to traverse all modes of the tensor with a major loop \cite{sthosvd-algo}.
Inside the loop, two important steps are carried out to compute the factor matrices $\bm{U}^{(n)}$ (line 5-6)
and update the core tensor $\bm{\mathcal{Y}}$ (line 8).
In the original $st$-HOSVD algorithm, the solver for the factor matrices and core tensor is based on
singular value decomposition (SVD) of the matricized input tensor.
Equivalently, this solver can be replaced with the eigen-decomposition of the Gram matrix of the matricized
input tensor \cite{ipdps16-tuckermpi} or with an alternating least squares (ALS) based iterative method \cite{xiaoals}.
In what follows, we call the three algorithm variants $st$-HOSVD-SVD, $st$-HOSVD-EIG, and $st$-HOSVD-ALS,
respectively.

\begin{figure}[!]
\centering
\includegraphics[width=0.9\columnwidth]{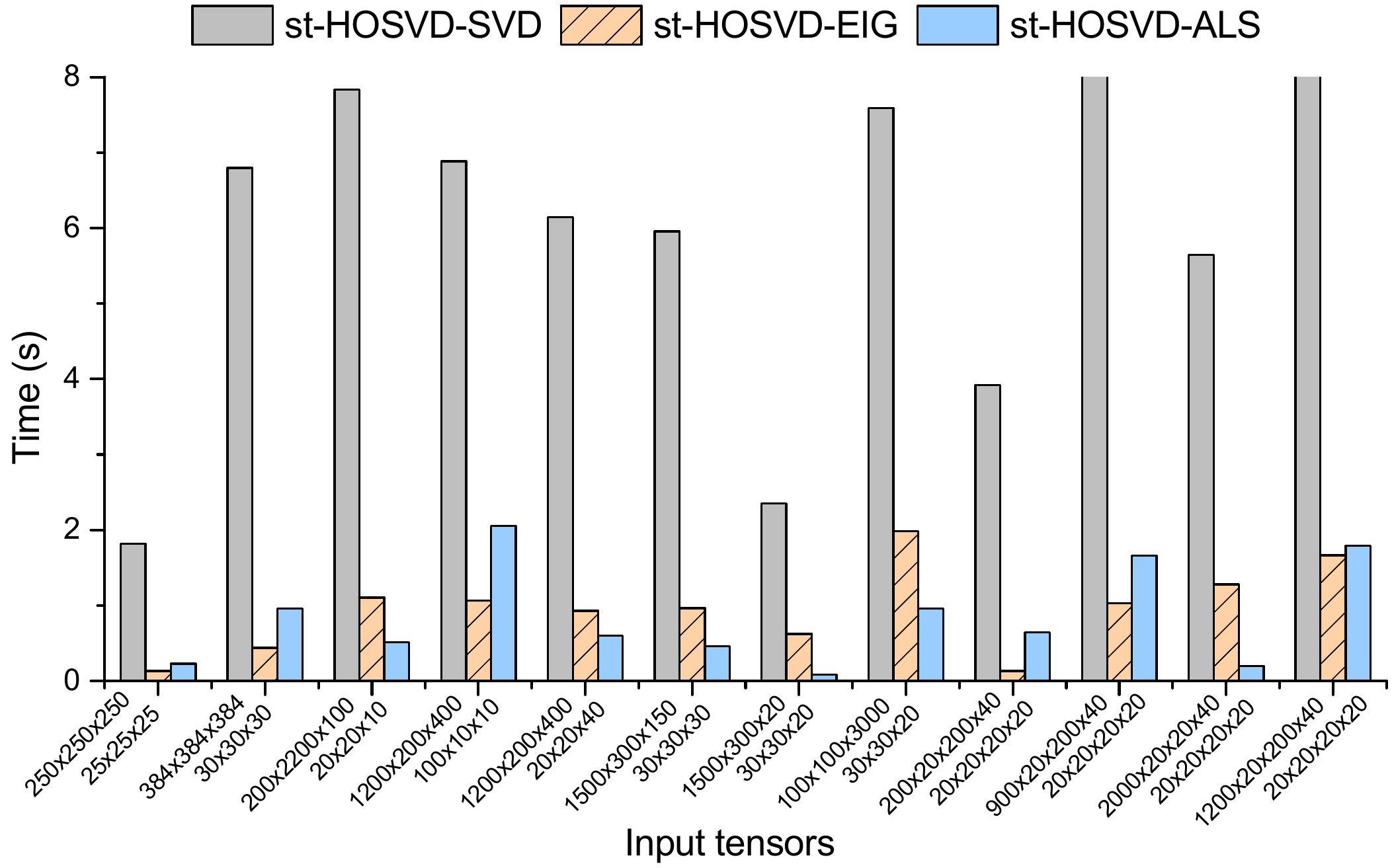}
\caption{The performance of the three $st$-HOSVD algorithm variants
for a number of synthetic tensors with different dimensions and truncations.}
\label{fig-motivation}
\end{figure}

In many applications, the dimensions of the input tensor and the truncations for Tucker decomposition can be diverse
and can usually spread in a large range.
To examine the performance of the three $st$-HOSVD algorithm variants, we perform
some experiments on {an Intel Xeon E5-2620 v4 CPU} with a number of synthetic tensors of different dimensions and truncations.
From the test results shown in Figure \ref{fig-motivation}, it can be seen that
despite the fact that the SVD based algorithm is the slowest in all tested cases,
it is hard to tell {\it a prior} whether the solver based on eigen-decomposition or that based on ALS is superior.
In fact,  the dimensions and truncations of the input tensors as well as the characteristics of
the hardware,  can all have a strong effect on the sustained performance.
It is therefore necessary to introduce more flexibility and adaptivity in the $st$-HOSVD algorithm.

\section{Flexible $st$-HOSVD Algorithm}
\label{algo}

In this section, we propose a new highly flexible $st$-HOSVD algorithm to cope with
various application scenarios, and present some formal cost analysis on it to further guide
the design of the adaptive solver selector and the subsequent performance optimizations.

\subsection{Mode-wise flexible algorithm}

A natural way to make the $st$-HOSVD algorithm more flexible is to enable the switch between
the eigen-decomposition and the ALS based methods, either of which is treated as a whole
throughout the major loop. Here it is not necessary to consider the original SVD based approach
due to its relatively inferior performance as tested in the previous section.
With further assistant of some automatic selecting mechanism,
it is possible to make the algorithm adapt with the change of the inputs and the hardware.
However, the question here is: is this flexible enough? The answer is of course no.
The key reason is that this coarse-grained flexible algorithm only has two possible choices,
greatly limiting its ability of coping with various application scenarios.
It is therefore necessary to introduce more degree of flexibility.

Instead of using a single solver for the factor matrices and core tensor across all modes,
we propose to orchestrate the major loop of the original $st$-HOSVD algorithm
and introduce the possibility of using different solvers for different tensor modes.
Algorithm \ref{algo-st-hosvd-fine-adaptive} demonstrates the proposed flexible $st$-HOSVD algorithm.
In the major loop, the flexible algorithm is able to make a dynamic switch between solvers based
on eigen-decomposition (line 6-8) and ALS (line 10-13).
The specific procedure of the ALS iteration is further shown in Algorithm \ref{algo-als}.
By switching the solvers for different modes, the flexible algorithm
can extend the solver selection space with the degree of flexibility increased
from 2 to $2^N$, where $N$ is the number of modes of the input tensor.
With the exponential extension of the selection space, it is more likely to achieve better performance
in the proposed flexible $st$-HOSVD algorithm.
It is worth mentioning that the accuracy of the proposed flexible algorithm
is maintained to a similar level of the $st$-HOSVD-EIG or $st$-HOSVD-ALS algorithm, as will be shown in later experiments.

\begin{algorithm}
\label{algo-st-hosvd-fine-adaptive}
\caption{Mode-wise flexible $st$-HOSVD algorithm.}
\SetAlgoLined
\KwIn{Input Tensor $\bm{\mathcal{X}}\in\mathbb{R}^{I_{1}\times I_{2}\times\cdots\times I_{N}}$, \\
      Core tensor dimensionality $(R_{1},R_{2},\cdots, R_{N})$  }
\KwOut{Core tensor $\bm{\mathcal{G}}\in\mathbb{R}^{R_{1}\times R_{2}\times\cdots\times R_{N}}$, \\
       Factor matrices $\bm{U}^{(n)}\in\mathbb{R}^{I_{n}\times R_{n}}$ $(n=1,2,\cdots,N)$  }
$\bm{\mathcal{Y}} \leftarrow \bm{\mathcal{X}} $ \\
 \For{$n \leftarrow 1 \ \KwTo \ N $} {
    method = algorithmSelector() \\
    Mode-$n$ matricization $\bm{Y}_{(n)} \leftarrow  \bm{\mathcal{Y}} $ \\
  \eIf{\rm{method} = 0}{
      $\bm{S} \leftarrow \bm{Y}_{(n)} \bm{Y}_{(n)}^{T} $ \\
      $\bm{U}^{(n)} \leftarrow R_{n}$ leading eigenvectors of $\bm{S} $ \\
      $\bm{Y}_{(n)} \leftarrow$ TTM$(\bm{\mathcal{Y}}, \bm{U}^{(n)T} ) $ \\
   }{
      $\bm{L, \mathcal{R}}  \leftarrow \mathrm{ALS} (\bm{Y}_{(n)}, R_n) $ \\
      $ \hat{\bm{Q}}, \hat{\bm{R}} \leftarrow $ QR decomposition on $\bm{L}$ \\
      $\bm{U}^{(n)} \leftarrow \hat{\bm{Q}} $ \\
      $\bm{Y}_{(n)} \leftarrow$ TTM$(\bm{\mathcal{R}}, \hat{\bm{R}}) $ \\
   }
   Tensorization $\bm{\mathcal{Y}} \leftarrow  \bm{Y}_{(n)} $
  }
  $\bm{\mathcal{G}} \leftarrow \bm{\mathcal{Y}}$
\end{algorithm}

\begin{algorithm}
\label{algo-als}
\caption{ $[\bm{L, \bm{\mathcal{R}}}]  = \mathrm{ALS} (\bm{Y}_{(n)}, R_n) $}
\SetAlgoLined
 \KwIn{Mode-$n$ matrix $\bm{Y}_{(n)}\in\mathbb{R}^{I_{n}\times J_n }$
       $(J_n=I_{1} \cdots  I_{n-1}  I_{n+1} \cdots  I_{N}),$ \\
       Truncation value $R_n,$ \\
      Initial guesses $\bm{L}_{0} \in\mathbb{R}^{I_{n}\times R_n}$ }
\KwOut{ $\bm{L, \mathcal{R}}$   }
 $k = 0$ \\
\While{not convergent}{
$\bm{{R}}_{k} \leftarrow (\bm{Y}_{(n)}^{T}\bm{L}_{k})(\bm{L}_{k}^{T}\bm{L}_{k})^{-1}$ \\
$\bm{L}_{k+1} \leftarrow (\bm{Y}_{(n)}\bm{{R}}_{k})(\bm{{R}}_{k}^{T}\bm{{R}}_{k})^{-1}$ \\
$k \leftarrow k+1$ \\
}
Tensorization $\bm{\mathcal{R}} \leftarrow  \bm{R}_{k} $
\end{algorithm}

\subsection{Cost analysis}
\label{analysis}
We analyze the cost of the flexible $st$-HOSVD algorithm, measured in total number of floating-point operations.
Here we use {$I_n$ and $R_n$ to represent the dimension of the tensor and truncation on mode-$n$,
and $J_n$ to represent the dimension products of all modes except mode-$n$.}

In the proposed flexible algorithm, the eigen-decomposition based solver consists of three operations,
i.e., Gram matrix computation, TTM, and eigen-decomposition.
For the Gram matrix computation, the number of elements in the resulting matrix is $I_n I_n / 2$,
and to get each resultant element, $2J_n$ floating-point computations are needed.
So the overall number of floating-point operations for Gram matrix computation is $I_n I_n J_n$.
For the TTM operation, the number of floating-point operations is $2 I_n R_n J_n$ with $R_n J_n$ resultant elements
and each of the elements obtained by $2I_n$ operations.
Furthermore, eigen-decomposition can be computed by calling routines from high performance LAPACK library.
Here, we use $f_{eig}(I_n,  I_n)$ to represent the cost of
the eigen-decomposition of an $I_n \times I_n$ matrix in terms of the number of floating-point operations.
Accordingly, the overall number of floating-point operations in the eigen-decomposition based solver
can be expressed as:

\begin{align}
F_{1}  =  &\underbrace{I_n  I_n  J_n}_{\texttt{Gram matrix computation}}
           + \underbrace{2 I_n R_n  J_n}_{\texttt{TTM computation}} \nonumber \\
           &+ \underbrace{f_{eig}(I_n,  I_n)}_{\texttt{Eigen-decomposition}}. \label{equation-4}
\end{align}

The ALS based solver, on the other hand, is comprised of multiple tensor/matrix operations and some other linear algebra operations.
In each inner iteration of the ALS procedure, two TTMs, two TTTs, two GEMMs and two matrix inversions are needed.
Outside the ALS iteration, one TTM and one QR decomposition are further utilized.
Therefore, the overall number of floating-point operations in the ALS based solver is:
\begin{align}
F_{2}  =  &(\underbrace{2 I_n J_n R_n + 2J_n R_n R_n}_{\texttt{TTM computation}}
           + \underbrace{2 I_n J_n R_n + 2J_n R_n R_n}_{\texttt{TTT computation}} \nonumber \\
           &+ \underbrace{4 I_n R_n R_n}_{\texttt{GEMM computation}}
           + \underbrace{2 f_{inv}(R_n, R_n)}_{\texttt{Matrix inversion}}) \times \rm{num\_iters}  \nonumber \\
           &+ \underbrace{2J_n R_n R_n}_{\texttt{TTM computation}} + \underbrace{f_{qr}(I_n, R_n)}_{\texttt{QR decomposition}}, \label{equation-5}
\end{align}
where $\rm{num\_iters}$ represents the number of ALS iterations that can be controlled by the users.
The default value is set to five in this work.
Analogously, $f_{qr}(I_n, R_n)$ and $f_{inv}(R_n, R_n)$ are used to represent the cost of the QR decomposition of an $I_n \times R_n$ matrix
and the cost of the matrix inversion of an $R_n \times R_n$ matrix, respectively.



\section{Adaptive Solver Selector}
\label{adapt}


The proposed flexible $st$-HOSVD algorithm relies on a selection mechanism
to pick up the most suitable solver for each mode.
This solver selector should be able to adapt with various inputs and hardware characteristics
and should be designed with low overhead and high accuracy.
As previously analyzed, the flexible $st$-HOSVD algorithm consists of a number of different operations.
It is therefore very difficult and complex to construct an accurate performance model for the solver selection.
To tackle this challenge, we utilize a lightweight machine learning model based on decision tree,
which can automatically learn the rules from the collected performance data.
Once the model is trained on a specific hardware platform, one can use it to make the prediction on-the-fly
as many times as necessary.

\subsection{Feature extraction}

\tabcolsep 1.6pt
\begin{table}[!h]
\scriptsize
\footnotesize
\renewcommand{\arraystretch}{1.3}
\caption{Extracted features for the decision tree based adaptive solver selector.}
\label{table:features}
\centering
\begin{tabular}{l|c}
 \hline
 Features & Meaning  \\
 \hline
    $I_n$         &dimension of the tensor in mode-$n$          \\
    $R_n$         &truncation in mode-$n$       \\
    $J_n$           &dimension products of all the modes except mode-$n$  \\
   \hline
   $I_n I_n$   &square of $I_n$                 \\
   $R_n R_n$   &square of $R_n$                 \\
   $I_n R_n$   &product of $I_n$ and $R_n$      \\
   $R_n R_n/I_n$  &ratio of square of $R_n$ and $I_n$   \\
   $R_n  R_n/J_n$   &ratio of square of $R_n$ and $J_n$      \\
   $I_n/J_n$        &ratio of $I_n$ and $J_n$      \\
   $R_n/J_n$        &ratio of $R_n$ and $J_n$      \\

   \hline
  \end{tabular}
\end{table}

Table~\ref{table:features} lists the extracted features for the decision tree model.
Firstly, basic shape information along mode-$n$ including the dimension size of the input tensor $I_n$, the truncation $R_n$,
and the dimension products of all proceeding modes $J_n$, is extracted.
{Then, we add some combined features derived from the three shape features.
Among them,  $I_n I_n$, $R_n R_n$ and $I_n R_n$ are used  to represent the computing scales of eigen-decomposition, QR decomposition and matrix inverse.
Information about the shape ratio is characterized by $I_n/J_n$ and $R_n/J_n$.
Furthermore, subtracting Equation (4) from Equation (5) and extracting the common factor $I_n J_n$,
another two features, i.e., $R_n R_n/I_n$ and $R_n  R_n/J_n$ are found in the major terms,
and are thus appended into the feature list.}

\subsection{Model training}

{To generate the decision tree model, a number of data samples, composed of features and the designated label,
are needed to be constructed. The statistics of each mode constitute a record in the sample database.
To obtain these records, third-order tensors are randomly generated as the inputs and
both the eigen-decomposition and ALS based solvers are tested for each mode.
The elapsed time is then recorded and used to serve as the basis of labeling.
The dimensions of the input tensors are set in the range of [10, 10000],
and the {truncation for each $n=1,2,\cdots,N$ is set in the range of $[10, 0.5 I_n]$,
which is a common configuration} in real applications.
Some input sizes are removed if they cannot fit into the main memory.
Overall, 1,500,000 samples are produced.
Then the data samples are divided into two parts with a split ratio of 7:3,
among which the former is used for training and the latter for testing.
The decision tree model is implemented based on the scikit-learn package \cite{sklearn}.
To tune the hyper-parameters, the grid search technique with cross-validation
is utilized to perform an exhaustive search over a range of parameters and find the best parameter set.
For the decision tree, we explore optimal parameters such as the maximum depth of the tree, i.e, max\_depth,
which is set in the range of [1,10], and the class weights, which belong to \{`balanced', `uniform'\}.
Once the model is trained, it can be converted to execution rules
and incorporated into the proposed flexible $st$-HOSVD algorithm
to help select the most appropriate method for each mode in an automatic way.

\begin{figure}[!b]
\centering
\includegraphics[width=0.95\columnwidth]{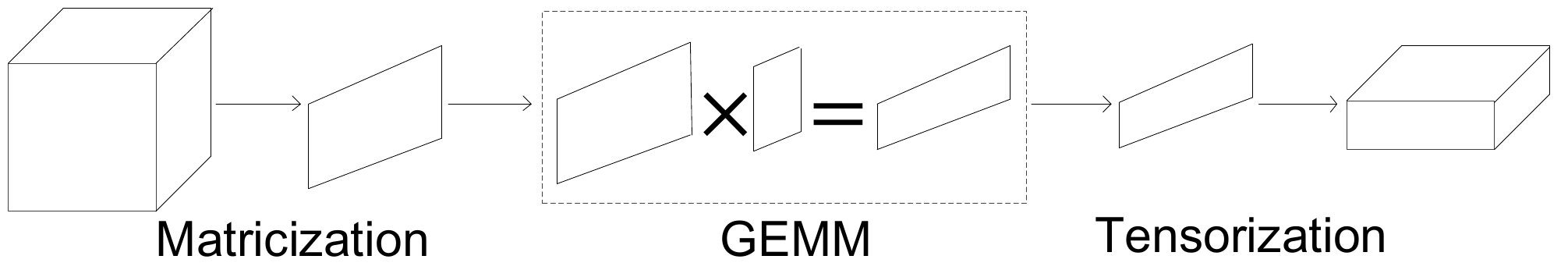}
\caption{The demonstration of explicit matricization method.}
\label{fig-explicit}
\end{figure}

\begin{figure*}[!]
\centering
\includegraphics[width=1.95\columnwidth]{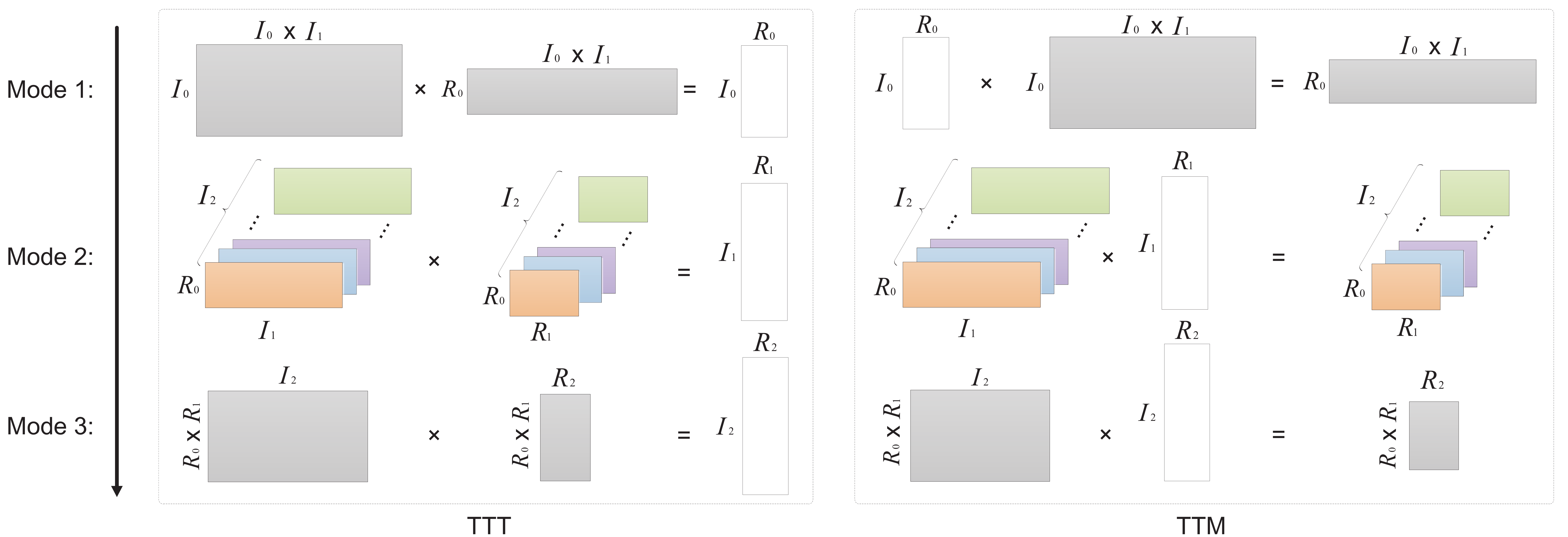}
\caption{Demonstration of the matricization-free method for TTT and TTM operations in the flexible $st$-HOSVD algorithm of a third-order tensor.
Here, the rectangles with color represent tensor, without color represent matrix.}
\label{fig-in-place}
\end{figure*}

\section{Implementation and Optimization}
\label{opt}

{To utilize the highly-optimized vendor supplied GEMM kernels,
many existing works on tensor computations \cite{tensortoolbox, tensorlab, xiaoals, sc18-dense-sthosvd-multi-gpu}
rely on explicit matricization techniques to improve the performance and simplify the implementation.}
Figure \ref{fig-explicit} demonstrates the typical workflow for tensor computing with the explicit matricization method, which usually consists three steps.
First, the input tensor is explicitly converted into a matrix by a mode-$n$ matricization operation along each mode.
Then the major computation is carried out by calling high-performance GEMM kernels.
After that, the resultant matrix is converted back to the tensor format through tensorization.
Despite the fact the whole procedure is easy to implement, the explicit matricization method
usually costs more memory for the storage of the intermediate matrices and brings extra conversion overheads.
To overcome this problem, we develop a matricization-free method for the proposed flexible $st$-HOSVD algorithm.

From the algorithmic perspective, the components of the flexible $st$-HOSVD algorithm are natural to be matricization-free,
because the major operations involving tensor are all tensor operations including Gramization, TTM, and TTT.
This fact indicates that if all theses tensor operations are implemented in a matricization-free way, so will the flexible algorithm be.
To achieve this goal, we consider all the involved tensor operations one by one.
Since the Gram matrix computation is a special case of TTT,
we only consider how to achieve matricization-free computations for TTM and TTT  operations.

As shown in Equation (2) and (3), the computation of TTM and TTT can be implemented by iterating the modes with nested ``for" loops.
In this case, to achieve high performance, two major factors need to be considered, i.e., loop ordering and loop merging.
In this work, the loop ordering is related to the underlying tensor storage and
we use column-major tensor layout, in which elements along the first dimension are stored contiguously.
Considering the locality of data access, the first dimension is more suitable to act as the most inner loop.
Thus, the loops are organized in the descending order, i.e., from the last dimension  of the tensor to the first.
To consider loop merging, it is known that two continuous dimensions of a tensor can be logically merged into one,
and different loop merging degrees can lead to different underlying kernel operations,
such as vector-vector, matrix-vector, or matrix-matrix operations.
To achieve high performance, matrix-matrix organization is usually more preferred.
Inspired by the explicit matricization of TTM and TTT on mode-$n$ which treats the $n$-th mode as a major dimension
for matrix-matrix multiplication, we use the $n$-th dimension as the splitting axis.
Then all the loops are split into three parts: loops outside, along, and inside the $n$-th axis.
By logically merging the loops outside and inside the $n$-th axis, the organization of the ``for" loops is simplified
and the kernel computation can be done by matrix-matrix operations straightforwardly.

Figure \ref{fig-in-place} demonstrates the matricization-free method
for TTT and TTM operations in the flexible $st$-HOSVD algorithm of a third-order tensor.
In the figure, the rectangles with color represent tensor, without color represent matrix.
As shown in the figure, there are no loops inside the first mode and no loops outside the last one,
and the computation is logically organized into a single GEMM operation naturally.
For the intermediate modes, i.e., the second mode, the computation is organized in the form of a series of GEMM operations.
The specific computation patterns of TTT and TTM are a little different.
For TTT, both inputs are multiple pieces of matrices, with the corresponding pieces multiplied one by one and written to a same resultant matrix.
As for TTM, the computation pattern is in fact the batched GEMM operation \cite{batchgemm}, whose performance can be improved
by calling the \texttt{batched\_gemm} kernels from Intel Math Kernel Library (MKL) \cite{mkl} and cuBLAS \cite{cublas} library.

\section{Performance Evaluation}
\label{experiment}

\tabcolsep 1.4pt
\begin{table}[!h]
\scriptsize
\footnotesize
\addtolength{\tabcolsep}{2.0pt}
\renewcommand{\arraystretch}{1.25}
\caption{Summary of real-world tensors used for experiments.}
\label{table:datasets}
\centering
\begin{tabular}{lllll}
 \hline
Name & Order &Dimension &Truncation &Abbr.\\
 \hline
   MNIST \cite{mnist}            &3   &784x5000x10     &65x142x10     &MNIST\\
    \hline
   Cavity\_velocity \cite{Burggraf1966} &3   &100x100x10000   &20x20x20      &Cavity\\
    \hline
   Boats \cite{boats-dataset}       &3    &320x240x7000        &10x10x10      &Boats\\
    \hline
   Air Quality \cite{tensor-data-set}     &3    &30648x376x6         &10x10x5      &Air\\
   \hline
   Sea-wave video \cite{sea-dataset}                 &4     &112x160x3x32        &10x10x3x32   &Video \\
   \hline
   HSI \cite{HSI-dataset}          &4     &1021x1340x33x8       &10x10x10x5    &HSI\\
   \hline
  \end{tabular}
\end{table}

\tabcolsep 3.2pt
\begin{table*}[!h]
\scriptsize
\footnotesize
\addtolength{\tabcolsep}{2.8pt}
\renewcommand{\arraystretch}{1.25}
\caption{The comparison of different $st$-HOSVD implementations for real-world tensors on both CPUs and GPUs.}
\label{table:perf}
\centering
\begin{tabular}{cl|cccccc|cccccc}
 \hline
 \multicolumn{2}{c|}{} & \multicolumn{6}{c|}{Approximation Error}  & \multicolumn{6}{c}{Time (Unit: s)}  \\  
 \hline
Platform & Implementation       &MNIST  &Cavity  &{Boats} &{Air} &Video &HSI &MNIST &Cavity  &Boats &Air  &Video &HSI\\
 \hline
 \multirow{3}*{CPU}
   &$st$-HOSVD-EIG    &0.213 &$0.00045$ &0.217 &0.291 &0.944&0.435 &10.90  &108.40  &34.80 &2804.80  &0.090 &{3.00} \\
   &{$st$-HOSVD-ALS}  &0.214 &$0.00045$ &0.219 &0.293 &0.945&0.442 &0.83   &2.09  &3.81 &0.43 &0.051 &{2.57} \\
   &a-Tucker          &0.214 &$0.00045$ &0.219 &0.293 &0.944 &0.443 & \textbf{0.48}   &\textbf{0.36} &\textbf{1.10} &\textbf{0.43} &\textbf{0.037} &\textbf{2.57} \\
   \hline\
   \multirow{3}*{GPU}
   &$st$-HOSVD-EIG    &0.213 &0.00045 &0.219 &-- &0.944 &0.435  &0.67          &3.75           &1.83                &--      &0.018                &0.30           \\
   &{$st$-HOSVD-ALS}  &0.214 &0.00045 &0.219 &0.292 &0.945 &0.436 &0.15      &1.43           &1.08               &0.24   &0.046                &0.21           \\
   &a-Tucker          &0.213 &0.00045 &0.219 &0.292 &0.944 &0.436 &\textbf{0.12}   &\textbf{0.43}  &\textbf{0.42}   &\textbf{0.24} &\textbf{0.018} &\textbf{0.21} \\
   \hline
  \end{tabular}
\end{table*}

\begin{figure*}[!]
\centering
\subfigure[On CPU]{\includegraphics[width=0.95\columnwidth]{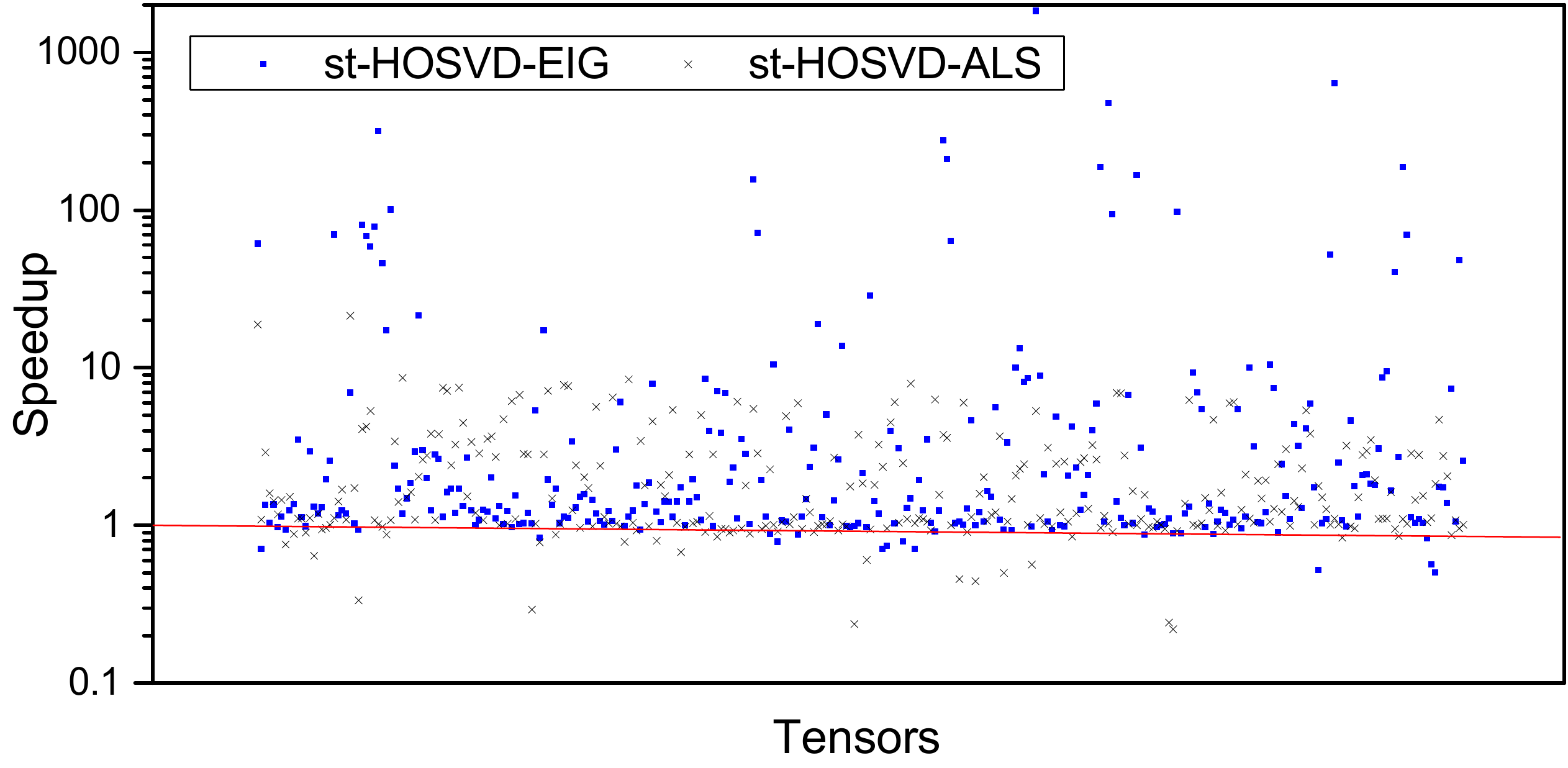}}
\subfigure[On GPU]{\includegraphics[width=0.95\columnwidth]{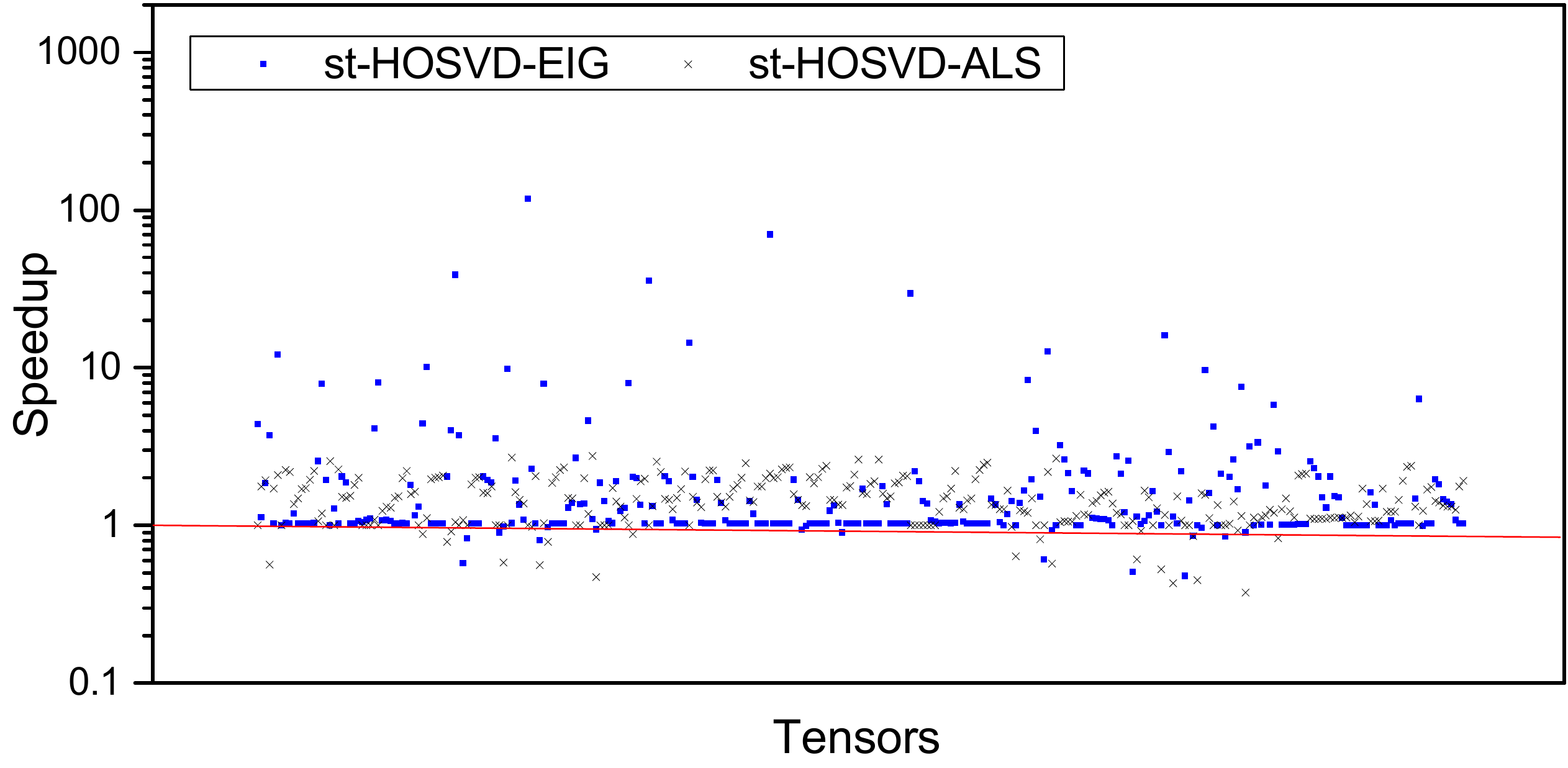}}
\caption{The achieved speedup of a-Tucker with respect to the $st$-HOSVD-EIG and $st$-HOSVD-ALS baselines.}
\label{fig-overall}
\end{figure*}

\subsection{Experiment setup}

\textbf{Platforms.} We ran all experiments on both CPU and GPU platforms.
The Intel Xeon CPU E5-2620 v4 @ 2.10GHz is used as the representative CPU platform,
which is with 32 cores and 64 GB of main memory.
As for GPU platform, the NVIDIA Tesla V100 (Volta) with 32 GB global memory is employed.
The CPU codes are compiled with g++ 5.4.0 and the GPU codes are compiled with nvcc 10.1,
both with \texttt{-O3} flag.
The double precision data type is used in all tests and the performance is measured
by averaging the results of five repeated executions.

\textbf{Datasets.} We use both synthetic tensors and tensors from real applications to evaluate the performance of a-Tucker.
A total of 300 synthetic tensors from the testing data sets,
introduced in Section \ref{adapt}, are used to examine the performance.
In addition to that, six tensors from real-world applications, covering both third-order and fourth-order tensors,
are tested for further performance analysis,
A list of the information of the six real-world tensors is shown in Table \ref{table:datasets},
where the entries ``Dimension" and ``Truncation" indicate the dimension of the input tensor and the truncation along each mode.
The specific values of the truncations are set to be consistent with existing works \cite{tensor-data-set,xiaoals}.

\textbf{Baselines.} This work is designed to compare with several other $st$-HOSVD implementations on both CPUs and GPUs.
The implementations based on eigen-decomposition and ALS alone are all utilized as the baselines.
On CPU, both single-node TuckerMPI library \cite{tuckermpi-software} and Tensor toolbox \cite{tensortoolbox}
provide eigen-decomposition based implementations.
Because Tensor toolbox is implemented by using MATLAB, to be more fair,
the single-node implementation in TuckerMPI library is selected as the baseline.
For ALS based method, we reproduce the $st$-HOSVD-ALS implementation according to \cite{xiaoals}.
{On the GPU platform, there are no readily available implementations for $st$-HOSVD.
We port the single-node TuckerMPI and $st$-HOSVD-ALS to GPU and take them as the reference implementations.}
{It is worth nothing that the proposed flexible algorithm and its implementation is in fact the breakdown and recombination of
the aforementioned baseline implementations of TuckerMPI and $st$-HOSVD-ALS.}
{For all the works, the involved linear-algebra operations are implemented by using the vendor supplied high-performance kernels
from the MKL \cite{mkl} library on CPUs, and the cuBLAS \cite{cublas} and cuSolver \cite{cusolver} libraries on GPUs.}

\subsection{Overall performance}

To compare the performance of a-Tucker and the baseline implementations, we test them with real-world tensors on both CPUs and GPUs.
Table \ref{table:perf} shows comparison results, including the approximation error and elapsed time.
Here, the approximation error ${\|\hat{\bm{\mathcal{X}}} - \bm{\mathcal{X}}\|_{F}}/{\|\bm{\mathcal{X}}\|_{F}}$
is used to evaluate the accuracy, where $\bm{\mathcal{X}}$ represents the input tensor and $\hat{\bm{\mathcal{X}}}$
the reconstruction from the output of Tucker decomposition.
From the table, it can be seen that a-Tucker can deliver results with accuracies similar to those of the reference $st$-HOSVD implementations, which further demonstrates its feasibility,
and at the same time, it can achieve better performance as compared to the baselines.
On the CPU platform, the achieved speedups are roughly 1.2x-6500.0x over $st$-HOSVD-EIG, and
1.0x-5.8x over $st$-HOSVD-ALS with the tested tensors.
Similar results can be found on GPUs, where a-Tucker can
yield 1.0x-8.7x and 1.0x-3.3x performance improvements over $st$-HOSVD-EIG and $st$-HOSVD-ALS, respectively.
{It is worth noting that for the Air Quality tensor, execution of $st$-STHOSVD-EIG is halted on GPU
because one of the dimensions of the tensor is too large to fit into the GPU memory space for eigen-decomposition of the Gram matrix.
a-Tucker, on the other hand, does not suffer from this memory issue.

To make more extensive performance comparisons, we use randomly generated tensors to do the tests on CPUs and GPUs again.
Figure \ref{fig-overall} shows the achieved speedup of a-Tucker with respect to the $st$-HOSVD-EIG and $st$-HOSVD-ALS baselines.
From the results, it can be seen that, among all 300 test cases,
a-Tucker can achieve better performance with around {91\%-94\%}  and 93\%-94\% cases on the CPU and GPU platforms, respectively.
On CPU, the average performance improvement is approximately {22.9x} and {2.2x}, with respect to $st$-HOSVD-EIG and $st$-HOSVD-ALS, respectively.
And on GPU, average speedups of {2.8x and 1.5x} are achieved when compared with $st$-HOSVD-EIG and $st$-HOSVD-ALS, respectively.
{It is also interesting to see that the achieved average speedup on CPU is larger than that on GPU.
This is caused by the reason that the performance gap of $st$-HOSVD-EIG and $st$-HOSVD-ALS on CPU is larger
than that on GPU due to the architectural differences.}
These test results clearly demonstrate the superior performance of a-Tucker.

\subsection{Flexibility and adaptivity}
\begin{figure}[!]
\centering
\subfigure{\includegraphics[width=0.94\columnwidth]{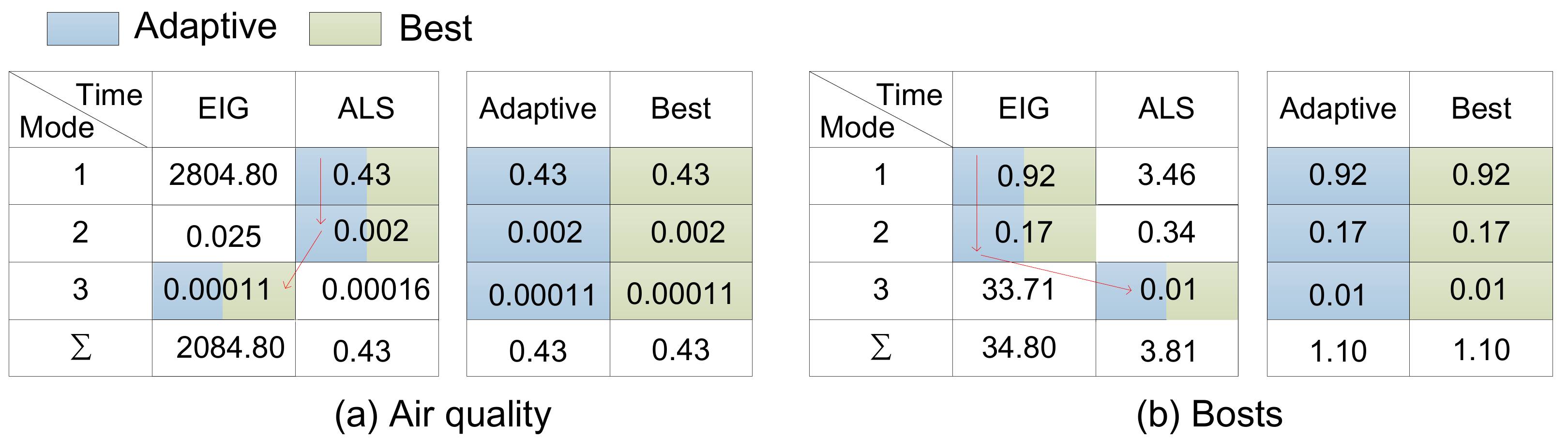}}
\caption{The runtime of different $st$-HOSVD algorithms with two real-world tensors on CPU.
"Adaptive" indicates the predicted solver and "Best" the true optimal one.}
\label{fig-runtime-example}
\end{figure}

To examine the effectiveness of the flexible and adaptive switch of the solver for the factor matrices and core tensor, the execution processes at the runtime of
the Tucker decomposition of two real-world tensors are provided in Figure \ref{fig-runtime-example}.
In the figure, the ``Adaptive" column indicates the predicted method by the adaptive solver selector,
and the ``Best" column indicates the true optimal method through exhaustive search.
The arrow points to the selected method for ``Adaptive" and ``Best" methods along the mode.
The CPU platform is used to show the testing results.
As shown in the figure, the flexible $st$-HOSVD algorithm can switch to the most suitable solver for each mode
with the help of the adaptive solver selector.
For some inputs, such as that in Figure \ref{fig-runtime-example} (a), the proposed mode-wise fine-grained flexible algorithm
achieves similar performance as the best alternative among $st$-HOSVD-EIG and $st$-HOSVD-ALS,
behaving similarly to the the coarse-grained adaptive algorithm.
For some other inputs, such as that in Figure \ref{fig-runtime-example} (b), the selection preferences in different modes
are quite different, which gives the proposed mode-wise algorithm more opportunity in flexibility,
leading to more performance improvement with the help of the adaptive solver selector.
These results further demonstrate the necessities and advantages of the proposed mode-wise flexible $st$-HOSVD algorithm
and the corresponding adaptive solver selector.

\subsection{Accuracy and overhead}
\begin{figure}[!b]
\centering
\includegraphics[width=0.72\columnwidth]{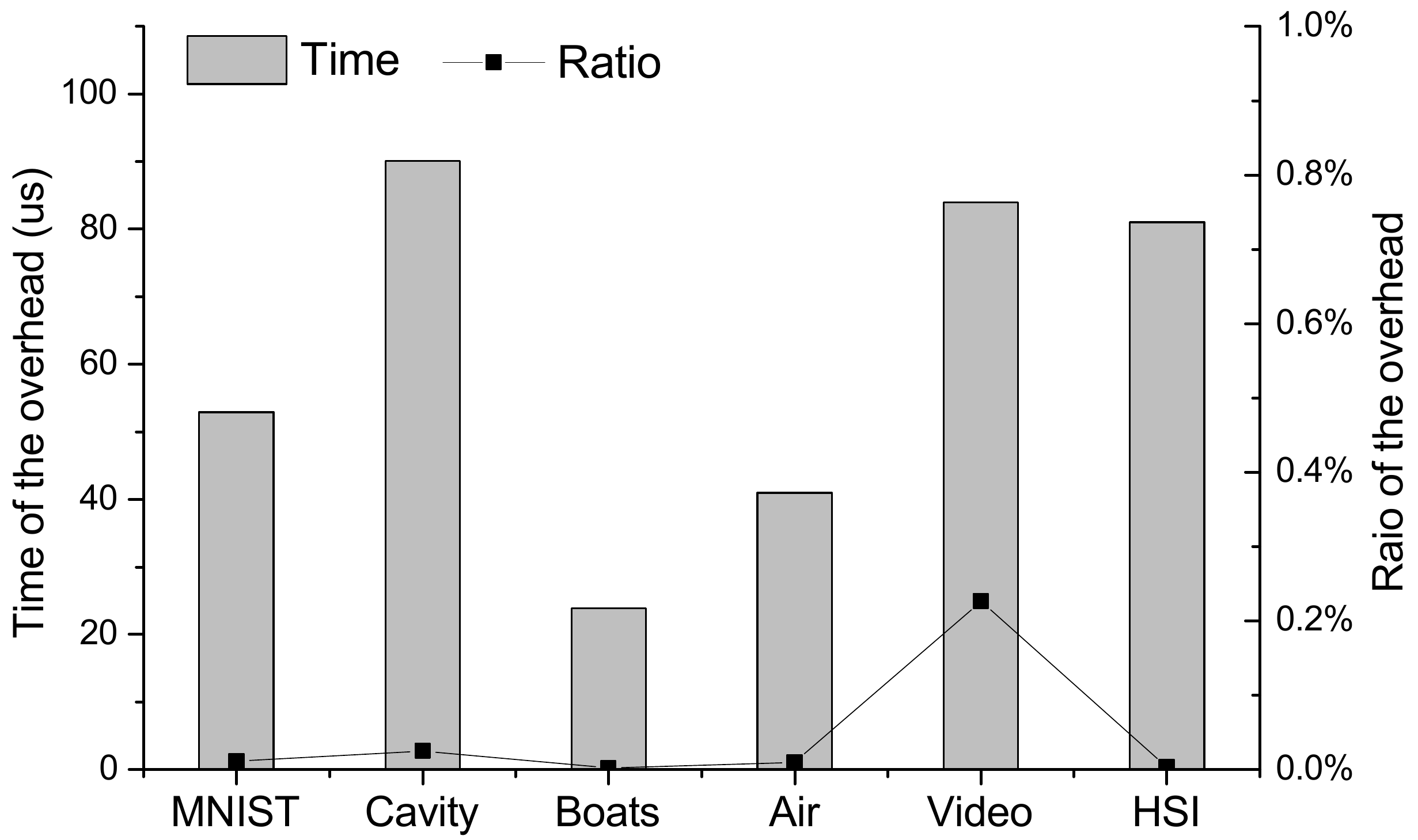}
\caption{The overhead of the adaptive solver selector. }
\label{fig-overhead}
\end{figure}

To investigate the accuracy of the machine-learning based adaptive solver selector,
we collect results on the prediction accuracy and find that the average prediction accuracy over all tested tensors
is around {92.9\%} and {93.7\%} on CPU and GPU platforms, respectively.
It is interesting to see that even for some wrongly predicted cases,
the flexible algorithm can still achieve better performance
due to its more flexible selection space, where the benefits from some modes
may compensate the extra cost from the wrongly predicted mode.

We further carry out test to analyze the runtime overhead of the adaptive solver selector.
Again, the CPU platform is utilized to analyze the results.
Figure \ref{fig-overhead} shows the overhead for dynamic switching between different solvers for
factor matrices and core tensor with the six real-world tensors.
As seen from the figure, the overhead is very small, which is only 23-90 us,
accounting to well below 0.25\% of the overall runtime,
even for small tensors such as the Sea-wave video tensor.

\begin{figure}[!]
\centering
\subfigure{\includegraphics[width=0.48\columnwidth]{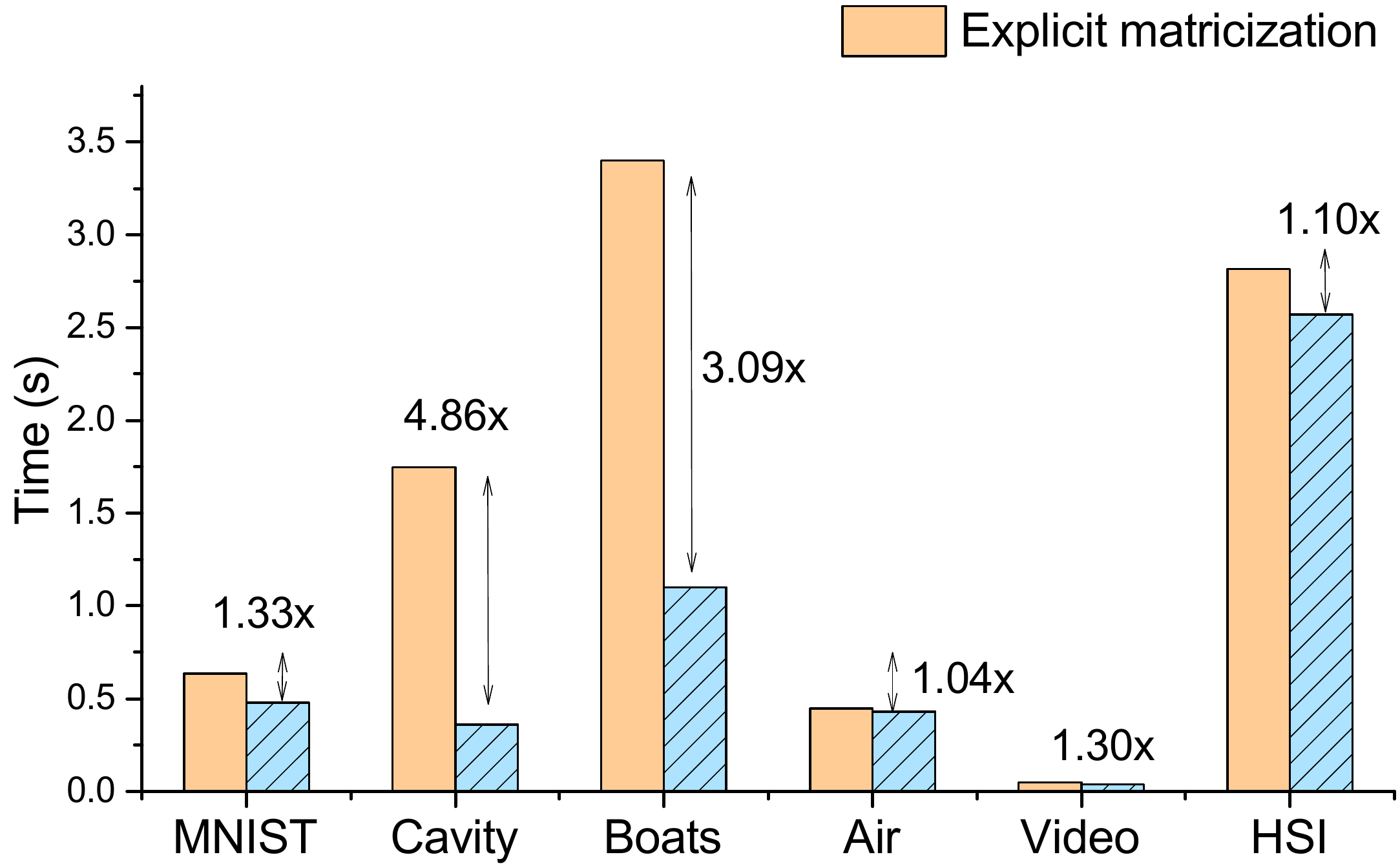}}
\subfigure{\includegraphics[width=0.48\columnwidth]{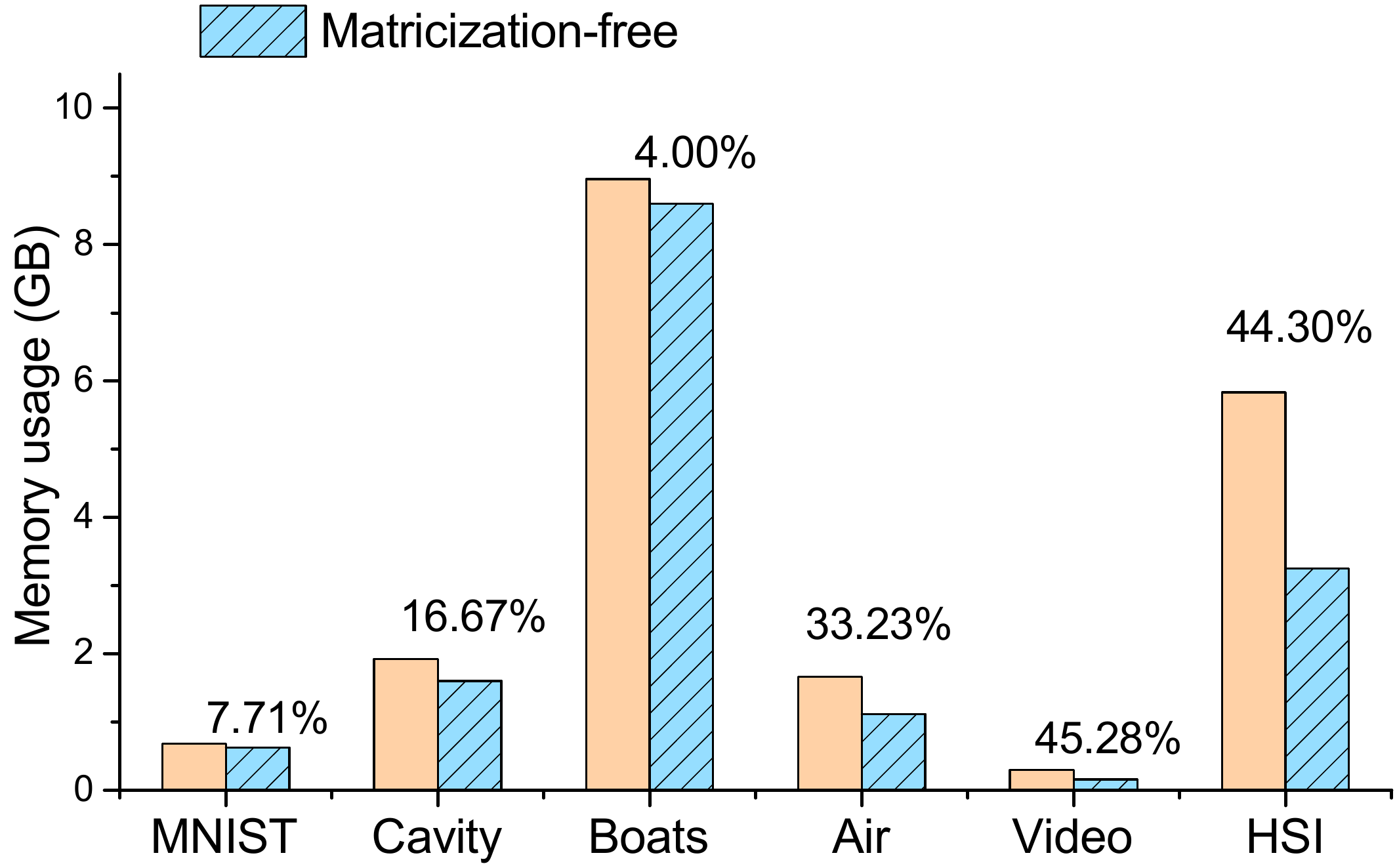}}
\caption{Performance improvement by using matricization-free optimization.}
\label{fig-optimization}
\end{figure}

\subsection{Matricization-free optimization}

To see how the matricization-free optimization can improve the overall performance of Tucker decomposition,
we conduct experiments on the proposed flexible $st$-HOSVD algorithm with the real-world tensors on the CPU platform,
using two implementations including the one with explicit matricization and the matricization-free approach.
We record both the execution time and the memory usage in Figure \ref{fig-optimization},
from which we can observe clearly that the matricization-free optimization can outperform the implementation with explicit matricization
with around 4\% to 386\% performance improvements, and at the same time reduce the memory space occupation
by around 4\% to 45\% for the tested tensors.
Overall, with the benefits from the elimination of explicit matricizaiton and tensorization operations,
the matricizaiton-free optimization has clear advantages in terms of both execution time and memory usage.


\section{Related Work}
\label{related-work}

In recent years, Tucker decomposition algorithms such as $st$-HOSVD are gaining increasingly more research attentions.
In addition to the original SVD based method for solving the factor matrices and core tensor in the $st$-HOSVD algorithm \cite{sthosvd-algo},
other approaches can be applied, such as the solver based on the eigen-decomposition of the Gram matrix \cite{ipdps16-tuckermpi}
and the iterative solver based on an ALS method \cite{xiaoals}.
Meanwhile, high performance {Tucker decomposition} libraries have been actively developed in the past decade.
Examples include Tensor toolbox \cite{tensortoolbox}, Tensorlab \cite{tensorlab} and TuckerMPI \cite{ipdps16-tuckermpi, tuckermpi-software}.
Efforts on performance optimizations of Tucker decomposition have also been done in various works,
focusing on both $st$-HOSVD \cite{sc18-dense-sthosvd-multi-gpu}
and HOOI \cite{hooi-distributed, s-hot, csf-europar2017}} algorithms.
However, most existing works only rely on a single solver for the factor matrices and core tensor.
In this paper, it is found that different solvers can play different roles inside the major loop of the algorithm.
To adapt with the diversities of the inputs and hardwares, we propose a highly flexible mode-wise algorithm
to enable the switch of different solvers for factor matrices and core tensor with different tensor modes.

Nowadays, with the rapid development of computer hardware, adaptive method for data format or kernel selection
has becoming a promising research direction, especially in applications that involve sparse matrix computations.
Examples can be found in works for automatic selection of sparse matrix format in SpMV
\cite{SMAT-spmv, 2016-icpp-spmv-select, spmv-model-likenli, adaptive-1, dnn-spmv, spmv-adaptive-1, spmv-adaptive-2}
and spGEMM \cite{sw-spMM, iaspgemm}, and the adaptive switch of computing kernels such as
SpMV/SpMSpV \cite{limin-spmspv} and SpTRSV \cite{ada-trsv}.
Recently, the adaptive idea is also used in MTTKRP sequence computation arising in the CP decomposition of higher-order tensors \cite{model-cp}.
In this paper, we have brought the idea of adaptivity into the Tucker decomposition and developed an adaptive computation framework.
With the help of a decision-tree based machine learning model, it can automatically select the optimal solvers for factor matrices and core tensor
with relatively low overhead and high accuracy, and can easily generalize to other hardware platforms.

There are also some works focusing on high performance optimizations of various basic tensor operations.
As one of the most important tensor operations in Tucker decomposition, TTM is well studied for
both dense tensors \cite{lijiajia-ttm} and sparse or semi-sparse tensors \cite{sparse-ttm-gpu}.
Besides, many efforts have made for the development of efficient tensor transpositions on CPUs and GPUs,
such as \cite{hptt, cutt, ttlg}.
There are also some works on the performance optimization of tensor-vector production \cite{sptv}
and tensor contraction, e.g., \cite{tensor-contraction-mawenjing2013, tensor-contraction-siam18, tensor-contraction-cgo19}.
In this paper, we have borrowed the matricization-free technique from reference \cite{lijiajia-ttm}
and applied it in both TTM and TTT operations of the flexible $st$-HOSVD algorithm.
Based on it, the Tucker decomposition has been successfully implemented in an adaptive and matricization-free manner on both CPUs and GPUs.


%


\section{Conclusion}
\label{conclusion}
The key finding of this paper is that the existing Tucker decomposition algorithms
are not flexible enough to cope with the diversities of the input data and the hardware.
To tackle this issue, a-Tucker, a new framework for input-adaptive and matricization-free Tucker decomposition of dense tensors, is proposed.
A mode-wise flexible Tucker decomposition algorithm is first proposed to enable the switch of
different solvers for factor matrices and core tensor with different tensor modes.
To automatically select the most appropriate solver at runtime,
a machine learning model is utilized to help the flexible algorithm adapt with different inputs
and also easily deploy on different hardware platforms.
To further improve the performance and reduce the memory usage, a matricization-free technique is employed on the implementation level.
Experiments show that a-Tucker can achieve substantially higher performance when compared with the previous state-of-the-arts
while keeping similar accuracy with both synthetic and real-world tensors.
Possible future work may include applying the proposed input-adaptive and matricization-free method to sparse tensors
and to other tensor decomposition models.

\bibliographystyle{IEEEtran}
\bibliography{STHOSVD}

\end{document}